\documentclass[onecolumn]{revtex4}

\usepackage{graphicx}
\usepackage{dcolumn}
\usepackage{bm}

\input epsf

\begin{document}

\title{\Large Scalar-Tensor Theory of Gravity and Generalized Second Law of Thermodynamics on the Event Horizon}

\author{\bf Nairwita~Mazumder\footnote{nairwita15@gmail.com},
Subenoy~Chakraborty\footnote{schakraborty@math.jdvu.ac.in}.}

\affiliation{$^1$Department of Mathematics,~Jadavpur
University,~Kolkata-32, India.}

\date{\today}

\begin{abstract}
In blackhole physics, the second law of thermodynamics is
generally valid whether the blackhole is a static or a non-static
one. Considering  the universe as a thermodynamical system  the
second law of blackhole dynamics extends to the non-negativity of
the sum of the entropy of the matter and the horizon , known as
generalized second law of thermodynamics(GSLT). Here, we have
assumed the universe to be bounded by the event-horizon or filled
with perfect fluid and holographic dark energy in two cases. Thus
considering entropy to be an arbitrary function of the area of the
event-horizon, we have tried to find the conditions and the
restrictions over the scalar field and equation of state for the
validity of the GSLT and both in quintessence-era and in
phantom-era in scalar tensor theory.
\end{abstract}


\maketitle

\section{\normalsize\bf{Introduction}}

In black hole physics, semi-classical description shows that just
like a black body , black hole emits thermal radiation (known as
Hawking radiation) and it completes the missing link between a
black hole and a thermodynamical system. The temperature (known as
the Hawking temperature) and the entropy (known as Bekenstein
entropy) are proportional to the surface gravity at the horizon
and area of the horizon [1,2] respectively (i.e. related to the
geometry of the horizon). Also this temperature , entropy and mass
of the black hole satisfy the first law of thermodynamics [3]. As
a result , physicists start speculating about the relationship
between the black hole thermodynamics and Einstein's field
equations (describing the geometry of space time). It is Jacobson
[4] who first derived Einstein field equations from the first law
of thermodynamics: $\delta Q=TdS $ for all local Rindler causal
horizons with $\delta Q$ and $T$ as the energy flux and Unruh
temperature seen by an accelerated  observer just inside the
horizon. Then Padmanabhan [5] was able to formulate the first law
of thermodynamics on the horizon, starting from Einstein equations
for a general static spherically symmetric space time.\\

Subsequently, this identity between Einstein equations and
thermodynamical laws has been applied in the cosmological context
considering universe as a thermodynamical system bounded by the
apparent  horizon ($R_A$). Using the Hawking temperature
$T_A=\frac{1}{2 \pi R_A}$ and Bekenstein entropy $S_A=\frac{\pi
R_A^2}G$ at the apparent horizon, the first law of thermodynamics
(on the apparent horizon) is shown to be equivalent to Friedmann
equations [6] and the generalized second law of thermodynamics
(GSLT) is obeyed at the horizon. Also in higher dimensional space
time the relation was established [6,7] for gravity with
Gauss-Bonnet
term and for the Lovelock gravity theory [8].\\

But difficulty arises if we consider universe to be bounded by
event horizon. First of all, in the usual standard big bang model
cosmological event horizon does not exists. However the
cosmological event horizon separates from that of the apparent
horizon only for the accelerating phase of the universe (dominated
by dark energy). Further, Wang et.al. [9] have shown that both
first and second law of thermodynamics break down at the event
horizon, considering the usual definition of temperature and
entropy as in the apparent horizon. According to them the
applicability of the first law of thermodynamics is restricted to
nearby states of local thermodynamic equilibrium while event
horizon reflects the global features of space time. Also due to
existence of the cosmological event horizon, the universe should
be non-static in nature and as a result the usual definition of
the thermodynamical quantities on the event horizon may not be as
simple as in the static space-time. They have considered the
universe bounded by the apparent horizon as a Bekenstein system as
Bekenstein 's entropy-mass bound: $S \leq 2E \pi R_A$ and
entropy-area bound:$S \leq \frac{A}4$ are valid in this region.
These Bekenstein bounds are universal in nature and all
gravitationally stable special regions with weak self gravity
satisfy Bekenstein bounds. Finally, they have argued that as event
horizon is larger than the apparent horizon  so the universe
bounded by the event horizon is not a Bekenstein system.

Recently Mazumder et al $[10,11]$ have examined the validity of
the GSLT on the event horizon assuming the validity of the first
law of thermodynamics on it. Without assuming any specific choice
of the entropy and temperature on the event horizon, they were
able to show the validity of the GSLT both for Einstein gravity
and for Gauss-Bonnet gravity. The restrictions on the matter in
the universe are the following :\\

(i) For flat and open FRW universe the generalized second law of
thermodynamics is valid if the weak energy condition $\rho+p>0$ is
satisfied.\\

(ii)For a closed model, the validity of the generalized second law
of thermodynamics demands either the weak energy condition is
satisfied and $R_{A}<R_H=\frac{1}{H}<R_{E}$ or the weak energy
condition is violated and $R_{A}<R_{E}<R_H$. \\

In this paper , we examine the validity of GSLT in Scalar-Tensor
theory when the universe is bounded by the event horizon is filled
with perfect fluid or holographic dark energy. The paper is
organized as follows: Section II deals with the basic equations in
Scalar-Tensor theory and its equivalence to Einstein gravity. The
first law of thermodynamics, Clausius relation and the time
variation of entropy functions are presented in section III . The
generalized second law of thermodynamics has been examined for
above two fluids and corresponding restrictions are determined in
section IV . The paper ends with a concluding remarks at the end
in section V.\\

\section{\normalsize\bf{Scalar-Tensor theory of gravity;}}

In Scalar-Tensor theory of gravity, using Jordan frame the
lagrangian is given by [12]
\begin{equation}
L=\frac{1}{16 \pi G}~f(\phi) ~R~-\frac{1}2~g^{\alpha \beta}~
\partial_{\alpha}\phi ~\partial_{\beta}\phi - V~(\phi)+L_m
\end{equation}

where $f(\phi)$ is a positive continuous otherwise arbitrary
function of the scalar field $\phi$ (having potential $V(\phi)$),
$L_m$ is the lagrangian for matter fields in the universe and R is
the Ricci curvature scalar of the space-time. For FRW model the
metric is

\begin{equation}
ds^{2}=-dt^{2}+a^2(t)\left[\frac{dr^2}{1-kr^2}+r^2{\Omega}_2^2\right]
\end{equation}

and we have $$R=6(\ddot{a}+{\dot{a}}^2)~.$$

Now varying the action corresponding to the Lagrangian (1) with
respect to the dynamical variables $g_{\mu \nu}$ and $\phi$ the
equation of motions are

\begin{equation}
G_{\alpha \beta}=\frac{8 \pi
G}{f(\phi)}\left[\partial_{\alpha}\phi~\partial_{\beta}\phi-\frac{1}2g_{\alpha
\beta}(g^{\mu \nu}\partial_{\mu}\phi \partial_{\nu}
\phi)-g_{\alpha \beta} V(\phi)-g_{\alpha \beta}\nabla^2 f+
\nabla_{\alpha} \nabla_{\beta} f +T_{\mu \nu}^{(m)}\right]
\end{equation}

and
\begin{equation}
\nabla^2 \phi - V'(\phi)+ \frac{1}{16 \pi G}f'(\phi) R=0~,
\end{equation}

where $T_{\mu \nu}^{(m)}$ is the energy-momentum tensor of the
matter fields.\\

If we assume $T_{\mu \nu}^{(m)}$ as the form of the
energy-momentum tensor for a perfect fluid i.e. $$T_{\alpha
\beta}^{(m)}=(\rho+p)U_{\alpha}U_{\beta}+p~g_{\alpha \beta}$$ with
$U^{\mu}$, the four velocity of the fluid, $\rho$ and $p$ the
energy density and the pressure of the fluid then the
non-vanishing components of the modified Einstein field equations
(3) and the scalar field equations (4) are

\begin{equation}
{H}^2+ \frac{k}{a^2}=\frac{8 \pi G}{3f}\left[ \rho + \frac{1}2
{\dot{\phi}}^2+V(\phi) - \frac{3H}{8 \pi G}f'\dot{\phi}\right]
\end{equation}

\begin{equation}
\dot{H}-\frac{k}{a^2}=-\frac{4 \pi
G}f\left[(\rho+p)+{\dot{\phi}}^2+\frac{1}{8 \pi
G}(f''{\dot{\phi}}^2+f'\ddot{\phi}-Hf'\dot{\phi})\right]
\end{equation}

and
\begin{equation}
\ddot{\phi}+3H\dot{\phi}+\frac{dV}{d \phi}=\frac{3}{8 \pi
G}({\dot{H}+H^2})f'
\end{equation}

or equivalently modified Friedmann equations can be written as

$$
{H}^2+ \frac{k}{a^2}=\frac{8 \pi G}{3f}\left[ \rho +
\rho_{ST}\right]=\frac{8 \pi G}{3f}
\rho_{eff}~~~~~~~~~~~~~~~~~~~~~~~~~~~~~~~~~~~~~~~~~~~~~~~~~~~~~~~~~~~~~~~~~~~~~~~~~~~~~~~~~~~~~~~~~~~~~(5a)
$$

$$\dot{H}-\frac{k}{a^2}=-\frac{4 \pi
G}f\left[(\rho+p)+(\rho_{ST}+p_{ST})\right]=-\frac{4 \pi
G}f(\rho_{eff}+p_{eff})~~~~~~~~~~~~~~~~~~~~~~~~~~~~~~~~~~~~~~~~~~~~~~~~~~~~~~~~~~~~~~~~~~(6a)$$

where
\begin{equation}
\rho_{ST}=\frac{1}2 {\dot{\phi}}^2+V(\phi) - \frac{3H}{8 \pi
G}f'\dot{\phi}
\end{equation}

\begin{equation}
p_{ST}=\frac{1}2{\dot{\phi}}^2-V(\phi)+\frac{1}{8 \pi
G}(f''{\dot{\phi}}^2+f'{\ddot{\phi}}+2Hf'\dot{\phi})
\end{equation}

and

\begin{equation}
\rho_{eff}=\rho+\rho_{ST}~,~p_{eff}=p+p_{ST}
\end{equation}

The energy conservation  relations are
\begin{equation}
\dot{\rho}+3H(\rho+p)=0
\end{equation}
and
\begin{equation}
{\dot{\rho}}_{ST}+3H(\rho_{ST}+p_{ST})=0
\end{equation}

Thus gravity in scalar-tensor theory is equivalent to Einstein
gravity with non-interacting two fluid system given by the field
equations (5a) and (6a).\\\\

\section{\normalsize\bf{Thermodynamic Study:}}

Let us start with FRW model for which the metric (2) can be
written as

\begin{equation}
ds^2=h_{ab}dx^adx^b+R^2d{\Omega}^2_2
\end{equation}

where $R=ar$ is the area radius and
$h_{ab}=diag\left(-1,\frac{a^2}{1-kr^2}\right)$
 with $ k=0,+1,-1$ for flat,closed and open model.\\

 Now to apply the first law of thermodynamics we define the work
 density W and the energy supply vector $\psi_a$ [13,14] as

 \begin{equation}
 W=-\frac{1}2T^{ab}h_{ab}
 \end{equation}
and
\begin{equation}
\psi_a=T_a^b \partial_b R +W\partial_a R.
\end{equation}

Here $T_{ab}$ is the projection of the four dimensional
energy-momentum tensor in the normal direction of the 2-sphere.
For universe bounded by the event horizon, $W$ gives the work done
due to a change of the event horizon while the total energy flow
through the event horizon is represented by the $\psi_a$. For FRW
model with perfect fluid as the matter content the expressions for
$W$ and $\psi_a$ are given by

\begin{equation}
W=\frac{1}2(\rho-p)~,~\psi_a=[-\frac{1}2(\rho+p)HR_E~,~\frac{1}2(\rho+p)a]
\end{equation}

where $R_E$ is the radius of the event horizon.\\

In thermodynamics , the clausius relation $\delta Q=TdS$ gives the
associated heat flow for a change in entropy. But heat flow to the
system is equivalent to a change of the energy of the system.
Hence the entropy of the event horizon is related to the energy
supply term by the relation

\begin{equation}
T_E dS_E=\delta Q=-dE=-A \psi=A(\rho+p)HR_E dt
\end{equation}

where$T_E~,~S_E$ are temperature and entropy of the event horizon
respectively and $A=4 \pi R_E^2$ is the area of the event
horizon.\\

For the present scalar-tensor theory (given by field equations
(5a) and (6a)) the above relation (17) becomes

\begin{equation}
\frac{dS_E}{dt}=\frac{4 \pi R_E^3 H}{fT_E}(\rho_{eff}+p_{eff})
\end{equation}

To find the change of entropy $S_I$ of the matter bounded by the
event horizon, we start with the Gibbs' equation [15]

\begin{equation}
T_EdS_I= dE_I+pdV
\end{equation}

where $E_I$ is the energy of the matter distribution , V is the
volume bounded by the event horizon and for the thermodynamical
equilibrium, the temperature of the matter is taken as that of the
event horizon i.e. $T_E$. Now using

\begin{equation}
V=\frac{4 \pi R_E^3}3~,~E_I=\frac{4}3 \pi R_E^3 ~\rho
\end{equation}

We have from the Gibbs' equation

\begin{equation}
\frac{dS_I}{dt}=\frac{4 \pi
R_E^2}{T_E}(\rho+p)\left[\frac{dR_E}{dt}-HR_E\right]
\end{equation}

\section{\normalsize\bf{Generalized Second Law of Thermodynamics:}}

We shall now examine the validity of the generalized second law of
thermodynamics (GSLT) for the following two cases:

\subsection{\bf{\underline{Universe filled with perfect fluid:}}}

In this case, the rate of change of the radius of the event
horizon is given by [10]$$\frac{dR_E}{dt}=HR_E-1$$

So from the equation (21)

$$
\frac{dS_I}{dt}=-\frac{4 \pi
R_E^2}{T_E}(\rho+p)~.~~~~~~~~~~~~~~~~~~~(21a)
$$

Thus combining equations (20) and (21a) we have

\begin{equation}
\frac{d}{dt}(S_E+S_I)=\frac{4 \pi R_E^2}{T_E}
\left[\left(\frac{HR_E}{f}-1\right)(\rho+p)+\frac{HR_E}{f}(\rho_{ST}+p_{ST})\right]
\end{equation}

Hence the validity of GSLT i.e. $\frac{1}{dt}(S_E+S_I)\geq 0$ we
have the following possibilities:\\\\

{\bf{I.}}  $~~~~~~\rho+p>0$, $R_E>fR_H$, $\rho_{ST}+p_{ST}>0$ i.e.\\

$~~~~~~~~\rho+p>0$ and $\frac{f}{R_E}<H<\frac{1}{f' \dot{\phi}}[
{\dot{\phi}}^2(8 \pi G+f'')+f'\ddot{\phi}]=H_{\phi}~$(say)\\

{\bf{II.}}$~~\rho+p>0$, $R_E<fR_H$, $\rho_{ST}+p_{ST}>0$ and
\begin{equation}
|\frac{\rho_{ST}+p_{ST}}{\rho+p}|>\frac{f R_H}{R_E}-1
\end{equation}

i.e. $\rho+p>0$, $H<~min[\frac{f }{R_E}~,~H_{\phi}]$ and
inequality
(23).\\

{\bf{III.}}$~~\rho+p>0$, $R_E>fR_H$, $\rho_{ST}+p_{ST}<0$ and
\begin{equation}
|\frac{\rho_{ST}+p_{ST}}{\rho+p}|<1-\frac{f R_H}{R_E}
\end{equation}

i.e. $\rho+p>0$, $H>~max[\frac{f }{R_E}~,~H_{\phi}]$ and
inequality (24).\\

{\bf{IV.}}$~~\rho+p<0$, $R_E<fR_H$ and $\rho_{ST}+p_{ST}>0$\\

i.e. $\rho+p<0$, $H<~min[\frac{f }{R_E}~,~H_{\phi}]$ and
inequality
(23).\\

{\bf{V.}}$~~\rho+p<0$, $R_E>fR_H$ , $\rho_{ST}+p_{ST}>0$ and inequality (23)\\

i.e. $\rho+p<0$, $\frac{f }{R_E}<H<~H_{\phi}$ and inequality
(23).\\

{\bf{VI.}}$~~\rho+p<0$, $R_E<fR_H$ , $\rho_{ST}+p_{ST}<0$ and inequality (24)\\

i.e. $\rho+p<0$, $ ~H_{\phi}<H<\frac{f }{R_E}$ and inequality
(24).\\

\subsection{\bf{\underline{Universe filled with holographic dark energy:}}}

Recent observational evidences demand that our universe is
experiencing an accelerated expansion driven by a missing energy
density with negative pressure (known as dark energy). An approach
to the problem of dark energy is holographic model[16-28]. The
holographic principle states that the no. of degrees of freedom
for a system within a finite region should be finite and is
bounded roughly by the area of its boundary. Using the effective
quantum field theory the energy density for holographic dark
energy can be written as[29]

$$\rho_{D}=3c^2{M_{p}}^2L^{-2} $$
where $L$ is an IR cut-off in units ${M_{p}}^{2}=1$ and $c~$is any
free dimensionless parameter whose value is determined by
observational data[23,30-34]. Li [16] has shown that $L=R_E$ gives
the correct equation of state and also the desired acceleration.
Thus we choose

\begin{equation}
\rho_{D}=\frac{3c^2}{R_E^2}
\end{equation}

Then from the definition of the cosmological event horizon

\begin{equation}
{R}_{E}=a\int^{\infty}_{a}\frac{da}{Ha^{2}}=\frac{c}{(\sqrt{\Omega_{D}})H}
\end{equation}

where ~$\Omega_{D}=\frac{\rho_{D}}{3H^{2}}$~is the density
parameter corresponding to dark energy.\\

Now using equation (25) and the energy conservation equation (11),
the time variation of $R_E$ has the expression

\begin{equation}
dR_{E}=\frac{3}2R_{E}H(1+\omega_{D})dt
\end{equation}

where $p_D=\omega_D\rho_D$ is the equation of state of the DE and
$\omega_D$ is not necessarily a constant. Using equation (27) in
(21) we get

\begin{equation}
\frac{d}{dt}(S_{I}+S_{E})=\frac{2\pi {R_{E}}^3H}{T_{E}}
(\rho_{D}+p_{D})(3\omega_{D}+1)
\end{equation}

Thus
\begin{equation}
\frac{d}{dt}(S_{I}+S_{E})=\frac{2\pi {R_{E}}^3H}{T_{E}}
\left[\rho_{D}(\omega_{D}+1)(\frac{2}f+3\omega_{D}+1)+\frac{2}f(\rho_{ST}+p_{ST})\right]
\end{equation}

Then for validity of GSLT any one of the following possibilities
must be satisfied.\\

{\bf{I.}}
$\frac{2}f+3\omega_{D}+1>0~,~\omega_{D}+1>0~~and~~\rho_{ST}+p_{ST}>0$
i.e.\\

$$\frac{-1}{2}<f<1~~if~~\omega_D<0~~~~~or~~~~~~~~f<\frac{-1}{2}~~if~~\omega_D<0$$
and $$H<H_{\phi}~~$$

{\bf{II.}}
$\frac{2}f+3\omega_{D}+1<0~,~\omega_{D}+1<0~~and~~\rho_{ST}+p_{ST}>0$
i.e.\\
$$\omega_D<\frac{2(1-f)}{3}-1~,~f>1~~and~~H<H_{\phi} $$

{\bf{III.}}
$\frac{2}f+3\omega_{D}+1>0~,~\omega_{D}+1>0~~and~~\rho_{ST}+p_{ST}<0$
and
\begin{equation}
|\frac{\rho_{ST}+p_{ST}}{\rho_D+p_D}|<|1+\frac{f
(3\omega_{D}+1)}{2}|
\end{equation}

i.e.

$$\frac{-1}{2}<f<1~~if~~\omega_D<0~~~~~or~~~~~~~~f<\frac{-1}{2}~~if~~\omega_D>0~,~
H>H_{\phi}~~and ~~the~~ inequality~(30)$$

{\bf{IV.}}
$\frac{2}f+3\omega_{D}+1<0~,~\omega_{D}+1<0~~and~~\rho_{ST}+p_{ST}<0$
 and inequality (30)  i.e.\\
$$\omega_D<\frac{2(1-f)}{3}-1~,~f>1~~and~~H>H_{\phi}~~and~~inequality~~(30) $$

{\bf{V.}}
$\frac{2}f+3\omega_{D}+1>0~,~\omega_{D}+1<0~~and~~\rho_{ST}+p_{ST}>0$
and
\begin{equation}
|\frac{\rho_{ST}+p_{ST}}{\rho_D+p_D}|>|1+\frac{f
(3\omega_{D}+1)}{2}|
\end{equation}

i.e.

$$\frac{3}2(1-f)-1<\omega_D<-1,~f>1,~
H<H_{\phi}~~and ~~inequality~~(31).$$

{\bf{IV.}}
$\frac{2}f+3\omega_{D}+1<0~,~\omega_{D}+1>0~~and~~\rho_{ST}+p_{ST}>0$
 and inequality (31)  i.e.\\
$$-1<\omega_D<\frac{3(1-f)}{2}-1~,~f<1~~and~~H<H_{\phi}~~and~~inequality~~(31) $$

\section{\normalsize\bf{Conclusions:}}

From the study of the validity of GSLT in Scalar-Tensor gravity
theory in the previous sections we have seen that as in other
gravity theory the results do not depend on the specific choice of
entropy function at the horizon- the only thing that we need is
the validity of the first law of thermodynamics at the event
horizon. For both the fluids, the geometric and physical
quantities are restricted one by the other for the validity of the
GSLT. For example, in case of perfect fluid the Hubble parameter
is restricted by geometry in one hand and by the scalar field on
the other. For dark energy model, both the equation of state
parameter and the Hubble parameter are restricted by the scalar
field for the validity of GSLT. At phantom divide line we have
$\rho+p=0$ (or $\rho_D+p_D=0$ in the case of DE) and GSLT will be
valid if  $\rho_{ST}+p_{ST}>0$ for both the matter. Finally, it is
interesting to note that if $f\rightarrow 1$ and $\phi \rightarrow
0$ then the conditions for the validity of GSLT at the event
horizon becomes identical as in Einstein gravity [10,11]. For
future work , it will be nice to infer about the entropy function
at the event horizon assuming the validity of GSLT there.\\

{\bf References:}\\
\\
$[1]$ S.W.Hawking, \it{Commun.Math.Phys} {\bf 43} 199 (1975).\\\\
$[2]$ J.D. Bekenstein, {\it Phys. Rev. D} {\bf 7} 2333 (1973).\\\\
$[3]$ J.M.Bardeen, B.Carter and S.W.Hawking, {\it
Commun.Math.Phys } {\bf 31} 161 (1973).\\\\
$[4]$ T.Jacobson, \it {Phys. Rev Lett.} {\bf 75} 1260 (1995). \\\\
$[5]$ T. Padmanavan , {\it Class. Quant. Grav. } {\bf 19} 5387
(2002).\\\\
$[6]$ R.G. Cai and S.P. Kim , {\it JHEP} {\bf 02} 050 (2005).\\\\
$[7]$ Akbar M. and Cai R.G. {\it Phys. Lett B} {\bf 635} 7
(2006)\\\\
$[8]$ Lancoz C. ,{\it Ann. Math.} {\bf 39 } 842 (1938)\\\\
$[9]$ B. Wang , Y. Gong , E. Abdalla, \it{Phys. Lett. B} {\bf 624}
141 (2005).\\\\
$[10]$ N. Mazumder and S. Chakraborty , {\it Class. Quant.
Gravity} {\bf 26} 195016 (2009).\\\\
$[11]$ N. Mazumder and S. Chakraborty , {\it Gen.Rel.Grav.} {\bf 42} 813
(2010).\\\\
$[12]$ R.G. Cai and M. Akbar, {\it Phys.Lett.B} {\bf 635} 7
(2006).\\\\
$[13]$ R.G. Cai and L.M. Cao , {\it Phys. Rev. D} {\bf 75} 064008
(2007).\\\\
$[14]$ R.G. Cai and L.M. Cao , {\it Nucl. Phys. B} {\bf 785} 135
(2007).\\\\
$[15]$ G. Izquierdo and D. Pavon, {\it Phys. Lett. B} {\bf 633}
420 (2006).\\\\
$[16]$ M. Li , {\it Phys. Lett. B} {\bf 603} 01 (2004);\\\\
$[17]$ M. R. Setare  and S. Shafei  , {\it JCAP } {\bf 0609} 011
(2006) arXiv: gr-qc/0606103 ;\\\\
$[18]$ B. Hu and Y. Ling \it{Phys. Rev. D} {\bf 73} 123510
(2006).\\\\
$[19]$ B. Wang , Y. Gong , E. Abdalla, \it{Phys. Lett. B} {\bf
624} 141 (2005).\\\\
$[20]$ M. Ito , {\it Europhys. Lett. } {\bf 71} 712 (2005);\\\\
$[21]$ S. Nojiri , S. Odintsov , {\it Gen. Rel. Grav.} {\bf 38}
1285(2006);\\\\
$[22]$ E.N. Saridakis , {\it Phys. Lett. B} {\bf 660} 138
(2008);\\\\
$[23]$ Q.G. Huang and M. Li , {\it JCAP} {\bf 0408} 013 (2004);\\\\
$[249]$ X. Zhang , {\it IJMPD } {\bf 14} 1597 (2005);\\\\
$[25]$ D. Pavon and W. Zimdahl , {\it Phys. Lett. B} {\bf 628} 206
(2005);\\\\
$[26]$ H. kim , H.W. Lee and Y.S. Myung , {\it Phys. Lett. B} {\bf
632} 605
(2006);\\\\
$[27]$ S.D. Hsu , {\it Phys. Lett. B} {\bf 594} 01 (2004);\\\\
$[28]$ R. Horvat , {\it Phys. Rev. D} {\bf 70} 087301 (2004);\\\\
$[29]$ A.G. Cohen , D.B. Kaplan and A.E. Nelson , {\it Phys. Rev.
Lett.} {\bf 82} 4971
(1999);\\\\
$[30]$ Z. Chang, F-Q. Wu and X. Zhang, Phys. Lett. B 633, 14
(2006) [arXiv:astroph/ 0509531].\\\\
$[31]$ H. C. Kao, W. L. Lee and F. L. Lin, astro-ph/0501487; X.
Zhang and F-Q. Wu, Phys. Rev. D 72, 043524 (2005)
[arXiv:astro-ph/0506310]; X. Zhang and F-Q. Wu, Phys. Rev. D 76,
023502 (2007) [arXiv:astro-ph/0701405].\\\\
$[32]$ Q. Wu, Y. Gong, A. Wang and J. S. Alcaniz,
[arXiv:astro-ph/0705.1006]; Y-Z. Ma and Y. Gong,
[arXiv:astro-ph/0711.1641].\\\\
$[33]$ J. Shen, B. Wang, E. Abdalla and R. K. Su,
[arXiv:hep-th/0412227].\\\\
$[34]$ E.N. Saridakis and M.R. Setare , {\it Phys. Lett. B} {\bf
670} 01 (2008);\\\\

\end{document}